\documentclass[letter]{ptptex}
\usepackage[dvipdfmx]{graphicx}
\usepackage{wrapft}

\notypesetlogo                       

\title{
Rescaled Perturbation Theory%
}

\author{
Tomoya \textsc{Hayata}\footnote{E-mail: t.hayata@nt.phys.s.u-tokyo.ac.jp}%
}

\inst{
Department of Physics, The University of Tokyo, Tokyo 113-0033, Japan
}

\abst{%
A nonperturbative method that can go beyond the weak coupling perturbation theory is introduced.
 The essential idea is to formulate a set of exact differential equations as a function of coupling strength $g$.
 Unlike other resummation methods in which information on higher-order terms is necessary,
 we only need a leading-order perturbative formula in every step to reach a large value of $g$.
 The method is applied to the quantum anharmonic oscillator and quantum double-well potential in one dimension.
 Both are known to have divergent series in the weak coupling perturbation and the latter is not Borel-summable.
 Our method is shown to work well from the weak coupling to the strong coupling for the energy eigenvalues and the wave functions.
 The method is also applied successfully to a system with a time-dependent external field.
}

\begin{document}

\maketitle

In quantum many-body problems and in quantum field theories, there are
 many systems where one needs to study phenomena
 over a broad range of coupling strengths \cite{Baym}.
 Typical examples are the BEC-BCS crossover in ultracold atoms where
 the coupling strength is changed by utilizing the Feschbach resonance
 \cite{Bloch,Ketterle},
 hadron-quark transitions in dense QCD (quantum choromodynamics)
 where the coupling runs with the baryon density \cite{Fukushima},
 and the quark-gluon plasma near the critical temperature where the
 coupling runs with temperature \cite{Andersen1}.
Thus far, a number of techniques have been developed to resum
 naive perturbative series or to produce a better convergent series
 under the names of resummed perturbation, optimized perturbation,
 variational perturbation and so on. (For reviews,
 see e.g., Refs. \citen{Arteca} and \citen{ZinnJustin:2010fb}.)

In this article, we introduce a novel method that can
 go beyond the weak coupling perturbation theory: we test its
 applicability to quantum mechanical examples such as the
 anharmonic oscillator (AHO) and double-well potential (DWP).
 Instead of resumming the perturbation series,
 we iterate the lowest order perturbation through the rescaling of the coupling strength,
 so that the method gives a global coupling dependence of eigenvalues and eigenvectors simultaneously.
 The basic formula has formal resemblance to the exact renormalization group method
 (ERG, see e.g., Ref. \citen{Sonoda} for a review) if we replace the cutoff scale of ERG
 by the coupling strength in the present approach.

Let us consider a quantum system such that the Hamiltonian is
\begin{equation}
\mathcal{H}(g)=\mathcal{H}_0+g\mathcal{H}_{\mathrm{int}}\;\;,
\end{equation}
where $g$ is a dimensionless coupling strength
 and $\mathcal{H}_0$ is assumed to be solved accurately.
 Our goal is to obtain $E_n(g)$ and $|\psi_n(g)\rangle$ satisfying 
\begin{equation}
\mathcal{H}(g)|\psi_n(g)\rangle=E_n(g)|\psi_n(g)\rangle\;\;.
\end{equation}
To simplify the discussion, we consider a nondegenerate $\mathcal{H}_0$.
 Generalization to the degenerate case is straightforward.
 Suppose we know $E_n(g)$ and $|\psi_n(g)\rangle$, then we can calculate
 $E_n(g+\delta{g})$ and $|\psi_n(g+\delta{g})\rangle$
 as long as $\delta{g}\ll{1}$ by using the leading-order
 Rayleigh-Schr\"{o}dinger perturbation theory (RSPT) for
 $\mathcal{H}(g+\delta{g})=\mathcal{H}(g)+\delta{g}\mathcal{H}_{\mathrm{int}}$,
\begin{align}
E_n(g+\delta{g})&=E_n(g)+\delta{g}{\langle}\psi_n(g)|\mathcal{H}_{\mathrm{int}}|\psi_n(g)\rangle \;\;,\\
|\psi_n(g+\delta{g})\rangle&=|\psi_n(g)\rangle+\delta{g}\sum_{n\neq{i}}\;\frac{\langle\psi_i(g)|\mathcal{H}_{\mathrm{int}}|\psi_n(g)\rangle}{E_n(g)-E_i(g)}|\psi_i(g)\rangle\;\;.
\end{align}

Then, we can set up differential equations for the eigenvalues,
 matrix elements $\mathrm{H}_{\mathrm{int}}^{ij}(g)={\langle}\psi_i(g)|\mathcal{H}_{\mathrm{int}}|\psi_j(g)\rangle$
 and the wave functions as
\begin{align}
\frac{d}{dg}E_i(g)&={\mathrm{H}}_{\mathrm{int}}^{ii}(g) \;\;,
\label{d1}\\
\frac{d}{dg}\mathrm{H}_{\mathrm{int}}^{ij}(g)&=
\sum_{i\neq{k}}\;\frac{\mathrm{H}_{\mathrm{int}}^{ik}(g)\mathrm{H}_{\mathrm{int}}^{kj}(g)}{E_i(g)-E_k(g)}
+\sum_{j\neq{k}}\;\frac{\mathrm{H}_{\mathrm{int}}^{ik}(g)\mathrm{H}_{\mathrm{int}}^{kj}(g)}{E_j(g)-E_k(g)} \;\;,
\label{d2}\\
\frac{d}{dg}|\psi_i(g)\rangle&=\sum_{i\neq{k}}\;\frac{\mathrm{H}_{\mathrm{int}}^{ki}(g)}{E_i(g)-E_k(g)}|\psi_k(g)\rangle\;\;.
\label{d3}
\end{align}
Note that we have no approximation to obtain the right-hand sides of
 these equations, so that they are exact equations.
 We take $\langle\psi_i(0)|\psi_i(g)\rangle=\langle\psi_i(0)|\psi_i(0)\rangle=1$
 as a normalization of the state vector, so that the norm of eigenvectors
 is conserved, i.e.,
 $\frac{d}{dg}\langle\psi_i(g)|\psi_i(g)\rangle=0$.
 Coupled differential equations (\ref{d1}), (\ref{d2}) and (\ref{d3})
 can be solved under the initial conditions, $E_i(0)$,
 $\mathrm{H}_{\mathrm{int}}^{ij}(0)$ and $|\psi_i(0)\rangle$
 and one may go to arbitrary large values of $g$ in principle.

Let us now apply this method to the one-dimensional quantum AHO:
\begin{equation}
\mathcal{H}_{\mathrm{AHO}}(g)=\frac{1}{2}p^2+\frac{1}{2}x^2+g{x^4}\;\;.
\label{AHO}
\end{equation}
This is a typical example that RSPT leads to a divergent series but is
 Borel-summable. (See e.g., Refs. \citen{Arteca} and \citen{Guillou} for reviews.)
 Using $c_{ij}(g)={\langle}\psi_j(0)|\psi_i(g)\rangle$, we can rewrite Eq. (\ref{d3}) as 
\begin{equation}
\frac{d}{dg}c_{ij}(g)=\sum_{i\neq{k}}\;\frac{\mathrm{H}_{\mathrm{int}}^{ki}(g)}{E_i(g)-E_k(g)}c_{kj}(g)\;\;.
\label{de3}
\end{equation}
The initial conditions at $g=0$ read,
\begin{align}
E_i(0)&=i+1/2 \;\;,\\
\mathrm{H}_{\mathrm{int}}^{ij}(0)&=\frac{1}{\sqrt{\pi2^{i+j}i!j!}}\int{dx}\;{e}^{-x^2}x^4 H_i(x) H_j(x) \;\;, \\
c_{ij}(0)&=\delta_{ij}\;\;,
\end{align}
where $H_i(x)$ is the $i$-th Hermite polynomial. In actual calculation,
 we solve the coupled differential equations for the states
 $i=0,1,\cdots{N-1}$. For $g$ up to 10, taking $N=50$ is accurate enough
 to have the eigenvalues of the low-lying states as shown in Table \ref{t2}.
 Here, we compare our results with those given in Ref. \citen{Biswas} by the diagonalization method.
\begin{table}[t]
\caption{Comparison of the lowest three eigenvalues of AHO.
 The upper numbers are obtained from the present method and the lower numbers from Ref. \citen{Biswas}.}%
\centering
\begin{tabular}{c|c|c|c}
\hline\hline
$g$ & $E_0$ & $E_1$ & $E_2$ \\ \hline 
$0.1$ & $0.55914633$ & $1.7695026$ & $3.1386243$ \\ 
      & $0.55914633$ & $1.7695026$ & $3.1386243$ \\ \hline
$0.5$ & $0.69617582$ & $2.3244064$ & $4.3275250$ \\
      & $0.69617582$ & $2.3244064$ & $4.3275250$ \\ \hline
$1.0$ & $0.80377065$ & $2.7378923$ & $5.1792917$ \\
      & $0.80377065$ & $2.7378923$ & $5.1792915$ \\ \hline
$5.0$ & $1.2245874$  & $4.2995081$ & $8.3179758$ \\
      & $1.2245870$  & $4.2995017$ & $8.3179605$ \\ \hline
$10$  & $1.5049814$  & $5.3216308$ & $10.348359$ \\
      & $1.5049724$  & $5.3216080$ & $10.347056$ \\ \hline
\end{tabular}%
\label{t2}
\end{table}

\begin{figure}[t]
\vspace{1ex}
\centering
\includegraphics[width=8cm]{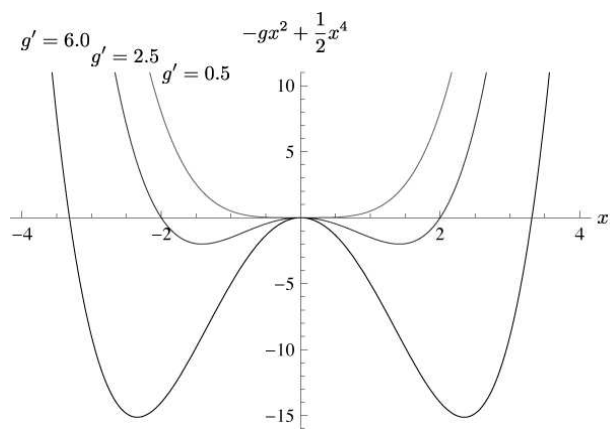}%
\caption{Double-well potential $V(x) = -gx^2 + \frac{1}{2}x^4$
 as a function of several values of $g'=g-1/2$.}%
\label{ff3}
\end{figure}%

Next, we study the quantum DWP with the Hamiltonian,
\begin{equation}
\mathcal{H}_{\mathrm{DWP}}(g)=\frac{1}{2}p^2-g{x^2}+\frac{1}{2}{x^4}\;\;,\;\;(g > 0)
\end{equation}
where the RSPT also breaks down and is known to be Borel-nonsummable.
 (See e.g., Refs. \citen{Arteca} and \citen{Guillou} for reviews.)
 To make a firm connection between DWP and AHO in our approach, we rewrite the Hamiltonian as
\begin{eqnarray}
\mathcal{H}^{\prime}_{\mathrm{DWP}}(g^{\prime}) &=& \Bigl(\frac{1}{2}p^2+\frac{1}{2}x^2+\frac{1}{2}{x^4}\Bigr)-g^{\prime}x^2 
\label{DWP}\\
&=&\mathcal{H}_{\mathrm{AHO}}-g^{\prime}\mathcal{H}_{\mathrm{int}} \;\;,
\end{eqnarray}
so that the unperturbed Hamiltonian at $g^{\prime}=0$ becomes AHO.
 Note that the system changes from
 AHO for $g' < 1/2$ to DWP for $g' > 1/2$.
 In Fig. \ref{ff3}, the shape of the potential is shown for several different values of $g'$.
 
 Using the results of the previous AHO for $N=50$,
 we calculate the energy eigenvalues of this system
 as a function of $g^{\prime}$.
 The differential equations are the same as Eqs. (\ref{d1}), (\ref{d2}) and (\ref{de3}),
 where $|\psi_i(0)\rangle$ is the eigenvector of AHO in the present case.
 The initial conditions read,
\begin{align}
E_i(0)&=E^{\mathrm{AHO}}_i\Bigl(\frac{1}{2}\Bigr) \;\;,\\
\mathrm{H}_{\mathrm{int}}^{ij}(0)&=\langle\psi_i(0)|-\hat{x}^2|\psi_j(0)\rangle \\
&=\sum_{k,l=0}^{N-1}c^{\mathrm{AHO}}_{ik}\Bigl(\frac{1}{2}\Bigr)c^{\mathrm{AHO}}_{jl}\Bigl(\frac{1}{2}\Bigr)
\frac{1}{\sqrt{\pi2^{k+l}k!l!}}\int{dx}\;{e}^{-x^2}(-x^2)
{H}_k(x){H}_l(x) \;\;, \notag \\
c_{ij}(0)&=\delta_{ij}\;\;.
\end{align}
Here, $i,j$ runs from $0$ through $N-1$, and $N$ is taken to be 50 as before.
 This is good enough again for low-lying eigenvalues as shown in Table \ref{t4}
 where comparison to the high-accuracy eigenvalues given in Ref. \citen{Balsa} by the diagonalization method is shown.
\begin{table}[t]
\caption{Comparison of the lowest two eigenvalues for the DWP.
 The upper numbers are calculated by our method, while the lower numbers are taken from Ref. \citen{Balsa}.}%
\centering
\begin{tabular}{c|c|cl}
\hline\hline
$g'$ & $E_0$ & $E_1$ \\  \hline 
$0.50$ & $0.53018104538$ & $1.8998365150$   \\
       & $0.53018104524$ & $1.8998365149$   \\ \hline
$1.0$  & $0.32882650295$ & $1.4172681012$   \\
       & $0.32882650260$ & $1.4172681011$   \\ \hline
$5.5$  & $-10.316788242$ & $-10.316773352$  \\
       & $-10.316788351$ & $-10.316773442$  \\ \hline
$8.0$  & $-25.420689499$ & $-25.420692377$  \\
       & $-25.420693642$ & $-25.420693642$  \\ \hline
\end{tabular}%
\label{t4}
\end{table}%

\begin{figure}[t]
\centering
\includegraphics[width=8cm]{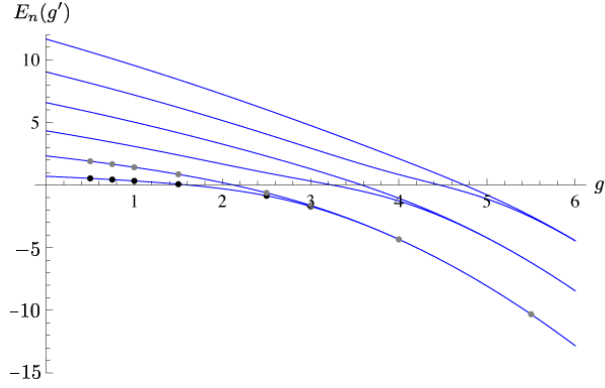}%
\caption{Lowest six energy eigenvalues for the DWP with $N=50$. 
Filled circles are high-accuracy numerical results given in Ref. \citen{Balsa}.}
\label{f1}
\end{figure}

\begin{figure}[t]
\vspace{1ex}
\centering
\includegraphics[width=8cm]{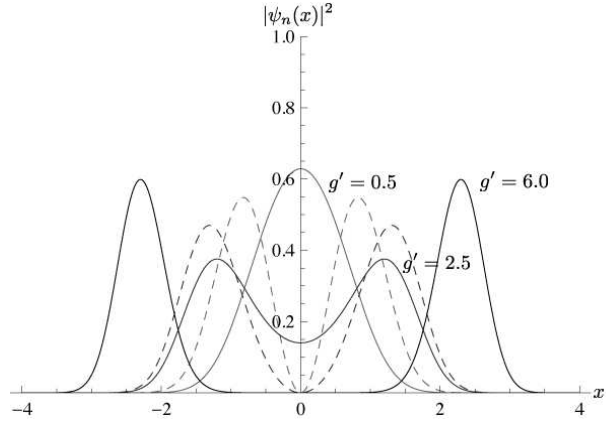}%
\caption{Probability distributions for the lowest two states for the DWP with $N=50$.
 The solid line is for the ground state and the dashed line is for the first excited state.
 The lighter gray line is for $g^{\prime}=0.5$, the gray line is for $g^{\prime}=2.5$ and
 the black line is for $g^{\prime}=6.0$.}%
\label{f2} 
\end{figure}

In Fig. \ref{f1},
 we show the $g'$ dependence of the lowest six energy eigenvalues
 from the weak coupling to the strong coupling.
 As $g'$ increases, the degeneracy of the even-$n$ and odd-$n$ eigenstates
 takes place starting from the low-energy state.
 This can be clearly seen by looking at the probability distributions of
 the lowest two states as shown in Fig. \ref{f2}.
 Depending on $g^{\prime}$, the energy eigenstate changes from
 a single oscillator to the superposition of
 double (left and right) oscillators.

The present method can be generalized to the time-dependent
 Schr\"{o}dinger equation with a nonadiabatic time-dependent potential:
\begin{equation}
i\partial_t|\psi(t)\rangle=\mathcal{H}(t)|\psi(t)\rangle\;,
\end{equation}
where $\mathcal{H}(t)=\mathcal{H}_0+g(t)\mathcal{H}_{\mathrm{int}}$.
 We expand the state vectors by the adiabatic basis:
\begin{align}
|\psi(t)\rangle&=\sum_n\;a_n(t)|n(t)\rangle \;, \\
\mathcal{H}(t)|n(t)\rangle&=E_n(t)|n(t)\rangle\;,
\end{align}
where $E_n(t)$ and $|n(t)\rangle$ are instantaneous eigenvalues
 and eigenvectors with their time dependence implicitly
 through the coupling $g(t)$.
 We perform a unitary transformation:
\begin{align}
a_n(t)&=\alpha_n(t){e}^{-i\Theta_n(t)} \;, \\
\Theta_n(t)&=\int_{0}^{t}dt^{\prime}\;E_n(t^{\prime})\;,
\end{align}
and obtain
\begin{equation}
\partial_t\alpha_n(t)=-\sum_m\;\alpha_m(t)\langle{n}(t)|\partial_t|m(t)\rangle{e}^{i(\Theta_n(t)-\Theta_m(t))}\;.
\end{equation}
Although $g(t)$ is an arbitrary smooth function of $t$,
 one may always introduce a set of time intervals, so that $g(t)$ is a monotonic function
 of $t$. Then, within each interval, we have
\begin{align}
\partial_{g}\alpha_n(g)&=-\sum_m\;\alpha_m(g)\langle{n}(g)|\partial_g|m(g)\rangle{e}^{i(\Theta_n(g)-\Theta_m(g))} \;, \\
\Theta_n(g)&=\int_0^{g}d\bar{g}\;\frac{1}{\dot{\bar{g}}}E_n(\bar{g})\;,
\end{align}
where $\dot{\bar{g}}$ implies the time derivative of coupling strength.
 Assuming that $\mathcal{H}_0$ is not degenerate, we obtain the following equations from (\ref{d1}), (\ref{d2}) and (\ref{d3}):
\begin{align}
\partial_{g}\alpha_n(g)&=
\sum_{m\neq{n}}\;\frac{\mathrm{H}_{\mathrm{int}}^{nm}(g)}{E_n(g)-E_m(g)}\alpha_m(g){e}^{i(\Theta_n(g)-\Theta_m(g))} \;,\\
\partial_{g}\Theta_n(g)&=\frac{1}{\dot{g}}E_n(g)\;.
\end{align}
To calculate the evolution of $\alpha_n(g)$ and $\Theta_n(g)$, we need the coupling dependence of $E_n(g)$ and
 $\mathrm{H}_{\mathrm{int}}^{nm}(g)$, which is directly obtained from our approach, so that we can trace the flows of $E_n(g)$,
 $\alpha_n(g)$ and $\Theta_n(g)$ simultaneously through the flows of $\mathrm{H}_{\mathrm{int}}^{nm}(g)$.

\begin{figure}[t]
\centering
\includegraphics[width=8cm]{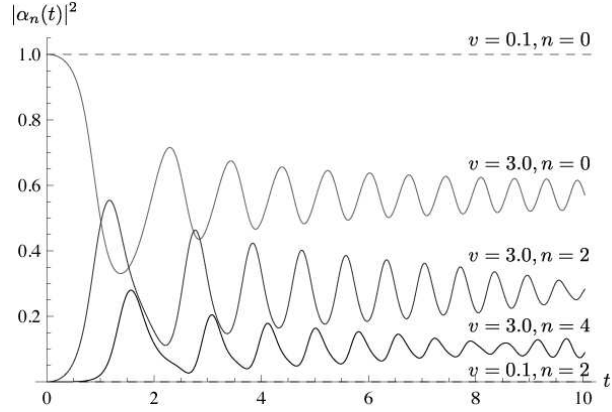}%
\caption{Lowest three even transition probabilities
 $|\alpha_{n}(t)|^2$ $(n=0,2,4)$ in DWP for $N=50$.
 The lighter gray line is for $n=0$, the gray line is for $n=2$ and
 the black line is for $n=4$.
 The dashed line is for $v=0.1$ and the solid line is for $v=3$.}%
\label{fff1}
\end{figure}

Here, we solve the differential equations numerically in the case that the coupling strength depends linearly
 on time (i.e., $g=vt$) and obtain the transition probabilities
 to adiabatic states $|\alpha_n(t)|^2$.
 In Fig. \ref{fff1}, taking the ground state of AHO as an initial condition,
 we show the transition probabilities as a function of time for the DWP with
 the Hamiltonian: $\mathcal{H}^{\prime}_{\mathrm{DWP}}(g^{\prime}(t)=vt)$ in Eq. (\ref{DWP}).
 Because of the parity conservation, the transition probability for odd $n$ becomes zero.
 If we change the potential slowly as the dashed lines ($v=0.1$) in Fig. \ref{fff1}, the
 instantaneous basis at $n=0$ is mostly occupied at any time.
 On the other hand, if we change the potential suddenly as the solid lines ($v=3.0$),
 higher instantaneous states are excited with mutual oscillations.

In summary, we introduced a nonperturbative method applicable to
 large values of the coupling strength $g$.
 The essential idea is to formulate a set of exact differential equations
 as a function of $g$, so that they can be solved under appropriate
 known initial conditions to reach large values of $g$. 
 Unlike resummation methods in which information on higher-order
 terms in the naive perturbation series is necessary, 
 we only need a leading-order perturbative formula in every step. This
 is similar to the situation in the exact renormalization group method.
 In the present method, all the eigenvalues and eigenstates can be calculated
 simultaneously and accurately as a function of $g$ as long as
 we prepare a sufficiently large enough dimension $N$ of the Hilbert space.
 Note here that we do not need to diagonalize the $N\times N$ Hamiltonian;
 instead, we trace the flows of eigenvalues and eigenvectors
 starting from the solvable $N\times N$ Hamiltonian.
 We have applied our method to the quantum anharmonic oscillator
 and quantum double-well potential in one dimension.
 Both are known to have divergent series in RSPT and the
 latter is not even Borel-summable. We found that our method works
 well from the weak coupling to the strong coupling for the
 energy eigenvalues and wave functions.
 Furthermore, because of the flow equations of the energy eigenvalues and the matrix elements
 $\mathrm{H}_{\mathrm{int}}^{ij}(g)$,
 we could solve even the time-dependent Schr\"{o}dinger equation for the potential with nonadiabatic variation in time.

 The basic idea of the present method can also be generalized to the
 quantum field theory and quantum statistical mechanics. In these cases,
 one can derive a differential equation for
 the partition function $Z[J,g]$ as functions of
 the external field $J$ and the coupling strength $g$,
\begin{equation}
\frac{d}{dg}Z[J,g]=\mathrm{S}_{\mathrm{int}}\Bigl(\frac{\delta}{\delta{J}}\Bigr)Z[J,g]\;, 
\end{equation}
where the initial condition $Z[J,0]$ is assumed to be solvable.
 Similarly, we can also apply the idea to the generalization of the linear response theory to derive a formula:
\begin{equation}
\frac{d}{dg}\langle\hat{O}\rangle[t;g] =
i\int_{t_0}^{t}dt^{\prime}\;\mathrm{Tr}\Bigl(\rho(t_0)[
\mathcal{H}^{h}_{\mathrm{int}}(t^{\prime};g),\hat{O}^{h}(t)]\Bigr)\;\;.
\end{equation}
Here, $A$ is an external field, $j$ is a conjugate quantity and $h$
 denotes a Heisenberg picture in terms of
 $\mathcal{H}[t;g]=\mathcal{H}_0+g\hat{j}A(t)$.
 Applications of these equations to physical systems will be
 discussed in forthcoming publications.

After the completion of this work, we became aware that 
 differential equations similar to ours are discussed in Appendix C of Ref. \citen{Kleinert},
 although the equations  are applied in a rather different context from ours.
 The author would like to thank Tetsuo Hatsuda, Kyogo Kawaguchi and Haruki Watanabe for useful discussions and suggestions.
%

\end{document}